\documentclass[twocolumn,showpacs,prl]{revtex4} 
\usepackage{graphicx}
\begin{document}  
\title{Competition between normal and intruder states inside the 
``Island of Inversion''} 
\author{Vandana Tripathi$^1$, S.L. Tabor$^1$, P.F. Mantica$^{2,3}$, 
Y. Utsuno$^4$, P. Bender$^1$, J. Cook$^2$, C.R. Hoffman$^1$,\\
Sangjin Lee$^1$, T. Otsuka$^{5,6}$, J. Pereira$^2$, M. Perry$^1$, 
K. Pepper$^1$, J. Pinter$^3$, J. Stoker$^3$,  A. Volya$^1$, D. Weisshaar$^2$}
\affiliation{\mbox {$^1$Department of Physics, Florida State University, 
Tallahassee, Florida 32306, USA} 
\mbox{$^2$National Superconducting Cyclotron Laboratory, Michigan State 
University, East Lansing, Michigan 48824, USA}
\mbox {$^3$Department of Chemistry, Michigan State University, East Lansing, 
Michigan 48824, USA}
\mbox{$^4$Japan Atomic Energy Agency, Tokai, Ibaraki 319-1195, Japan }
\mbox{$^5$Dept. of Physics and Center for Nuclear Study, University of Tokyo, 
Hongo, Tokyo 113-0033, Japan }
\mbox {$^6$RIKEN, Hirosawa, Wako-shi, Saitama 351-0198, Japan }
}
\date{\today}

\begin{abstract}

The $\beta^-$ decay of the exotic $^{30}$Ne ($N=20$) is reported.  For
the  first time,  the low-energy  level structure  of the  {\it N=19},
$^{30}$Na  (T$_Z$  = 4),  is  obtained  from $\beta$-delayed  $\gamma$
spectroscopy  using $fragment$-$\beta$-$\gamma$-$\gamma$ coincidences.
The  level  structure clearly  displays ``inversion'', {\it  i.e},
intruder  states  with mainly  $2p2h$  configurations displacing  the
normal  states to higher  excitation energies.  The good  agreement in
excitation  energies and  the weak  and electromagnetic  decay patterns
with Monte Carlo Shell  Model calculations with the SDPF-M interaction
in  the  $sdpf$  valence  space  illustrates  the  small  $d_{3/2}$  -
$f_{7/2}$  shell  gap.  The  relative  position  of  the  {\it  normal
dominant} and  {\it intruder dominant} excited states provides valuable 
information to understand better the $N=20$ shell gap.

\end{abstract}
\maketitle

The  anomalously large binding  energies of  neutron-rich $^{31,32}$Na
observed by Thibault {\it et  al.}, \cite {thibault} in 1975 offered a
tantalizing glimpse into a new era in nuclear structure physics -- one
which saw the collapse of  the conventional shell model.  The textbook
picture of fixed shell gaps and magic numbers was challenged as it was
realized that  the shell gaps could evolve, as  a result of
the shifting of  single particle levels in nuclei  with a large excess
of  neutrons,   due  to  the  spin-isospin  dependence   of  the  $NN$
interaction.   The  term  ``Island   of  Inversion''  was  applied  by
Warburton \cite {warburton}  to a region of nuclei  with $Z=10-12$ and
$N=20-22$ due  to their  tendency toward prolate  deformation despite
the  spherical driving  force of  the $N=20$  magic number.   Today we
understand this unexpected deformation  as a result of strong intruder
configurations in the ground states  of these nuclei, a consequence of
the reduced $N=20$ shell gap \cite {otsuka}.

Although there is a  consensus, both theoretically and experimentally,
about the inclusion of $fp$  configurations in the $N=20$ isotones for
$Z=10-12$,  the same  cannot  be said  about  the competition  between
$0p0h$ and $2p2h$  configurations for nuclei with $N<20$  or about the
degree  of mixing between  the various  configurations. Both these
depend critically on the $1d_{3/2}-f_{7/2}$  gap and to some extent on
the  $1f_{7/2}-p_{3/2}$   gap.   Different  nucleon-nucleon  effective
interactions used in current nuclear structure models give predictions
which  smear  the  `island  of inversion'  to  a  larger  or smaller
extent.  That used in the  Monte Carlo Shell Model (MCSM) calculations
by Utsuno  {\it et al.},  \cite {utsuno,utsuno2} creates  the smallest
$1d_{3/2}-fp$ gap as a function of  $Z$ (1.2 MeV for $^{28}$O to 4.4
MeV for $^{34}$S) and, thus,  an enlarged  `island of inversion'  and
enhanced  intruder mixing.   Only experiments  can select  between the
available models, a job rendered difficult due to the low luminosity of 
these exotic nuclei.

In  this Letter,  we  report  how a  detailed  spectroscopic study  of
$^{30}$Na ($N=19$) presents evidence for normal- and intruder-dominant
states at  low excitation energy and provides  the first comprehensive
look  at their  competition for  a Na  isotope inside  the  `island of
inversion', shown to  start at $N = 18$ for  Na \cite {tripathi}.  The
excited levels of  $^{30}$Na up to the neutron  separation energy were
populated following  the $\beta^-$  decay of $^{30}$Ne  ($N=20$).  The
selectivity  of  allowed  $\beta$  decay  from a  spin  $0^+$  nucleus
provides  firm  $J^\pi$  assignments.   The knowledge  gained  of  the
alteration in nuclear structure due to the large excess of one type of
particle   provides  an  excellent   opportunity  to   understand  the
isospin-dependent part  of the interaction in the  nuclear medium.  In
particular, the structure of $^{30}$Na, with an unpaired neutron close
to  $N  =  20$, is  predicted  to  be  particularly sensitive  to  the
$1d_{3/2}-f_{7/2}$  gap  and thus  provides  a  valuable  test of  the
effective interaction \cite{utsuno_na}.

 \begin{figure}
\begin{center}
\includegraphics[scale=0.35]{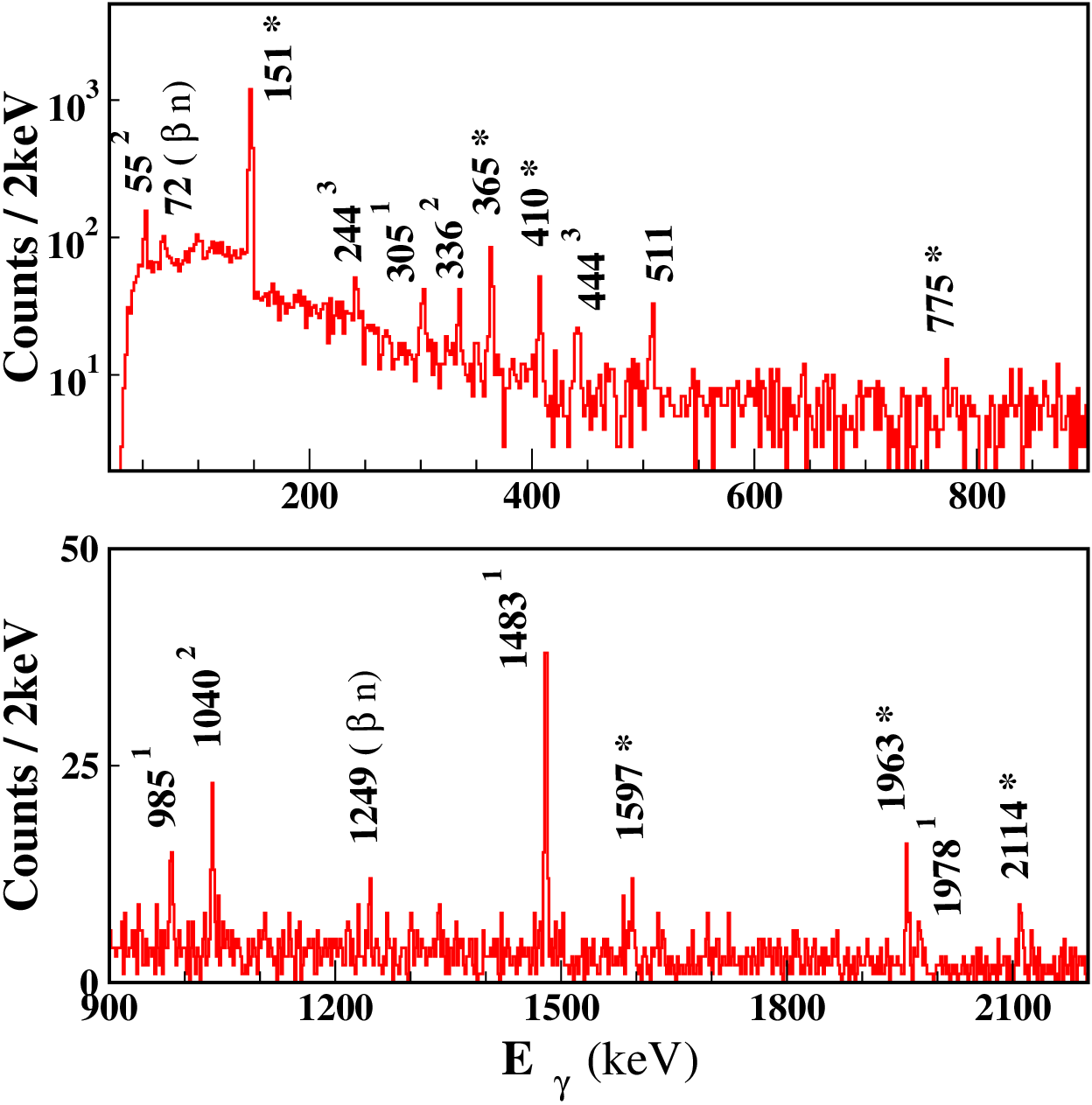}
\caption{(Color online) $\gamma$ spectrum  for events within the first
50  ms after a  $\beta^-$ correlated  $^{30}$Ne implant.  The $\gamma$
rays assigned to $^{30}$Na (asterisk) and transitions in the $\beta$-n
decay daughter, $^{29}$Na, are indicated.  Other transitions originate
from daughter and grand daughter activity, 1: $^{30}$Mg; 2: $^{29}$Mg;
3: $^{30}$Al.}
\label{fig1}
\end{center}
\end{figure} 

The  $\beta^-$ decay  of $^{30}$Ne  was investigated  at  the National
Superconducting   Cyclotron  Laboratory   (NSCL)  at   Michigan  State
University.  A 140 MeV/nucleon $^{48}$Ca$^{20+}$ beam of $\sim$ 75 pnA
was  fragmented in a  752 mg/cm$^2$  Be target  located at  the object
position  of  the  A1900  fragment  separator, used  to  disperse  the
fragments according  to their $A/Z$.  A 300  mg/cm$^2$ wedge-shaped Al
degrader placed  at the  intermediate image of  the A1900,  allowed to
separate  the transmitted  fragments  according to  $Z$. The  magnetic
rigidities of the A1900 magnets were set to 4.7856 Tm and 4.6558 Tm to
select the $^{30}$Ne  ions. With a momentum acceptance  of 2\% for the
A1900, the  yield of $^{30}$Ne was $\approx$  0.09 $s^{-1}pnA^{-1}$ at
the  Beta  Counting  System   (BCS)  \cite  {prisci}.   The  secondary
fragments  were unambiguously  identified by  a combination  of energy
loss and time-of-flight information and passed through a 2.6 mg/cm$^2$
Al  degrader  before implantation  in  the  40  x 40  Double-sided  Si
micro-Strip Detector  (DSSD). The DSSD, part  of the BCS,  was used to
detect both  the high-energy  fragments and the  subsequent low-energy
decay products.  Each  recorded event had a time  stamp generated by a
free running clock. The details of the experimental setup were similar
to  those   in  our  previous  investigation   of  $^{28,29}$Na  \cite
{tripathi,tripathi1},  except  that  16  detectors  of  the  Segmented
Germanium Array (SeGA) \cite{mueller}  were used instead of 12, giving
25\% higher $\gamma$-detection efficiency.

\begin{figure}
\begin{center}
\includegraphics[scale=0.35]{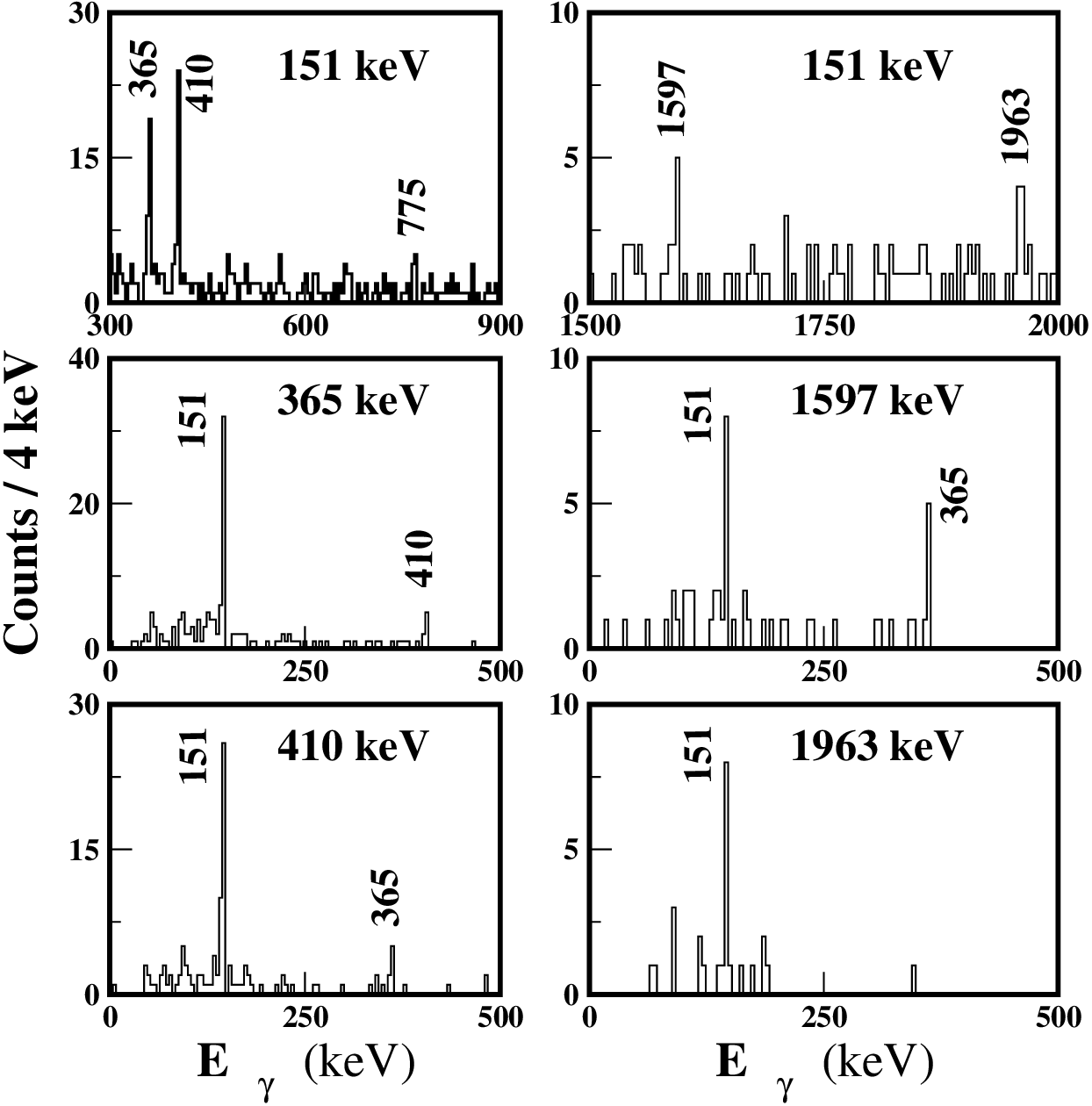}
\caption{$^{30}Ne-\beta-\gamma-\gamma$ coincidences, gating $\gamma$
indicated. }
\label{fig2}
\end{center}
\end{figure}

\begin{figure}
\begin{center}
\includegraphics[scale=0.38]{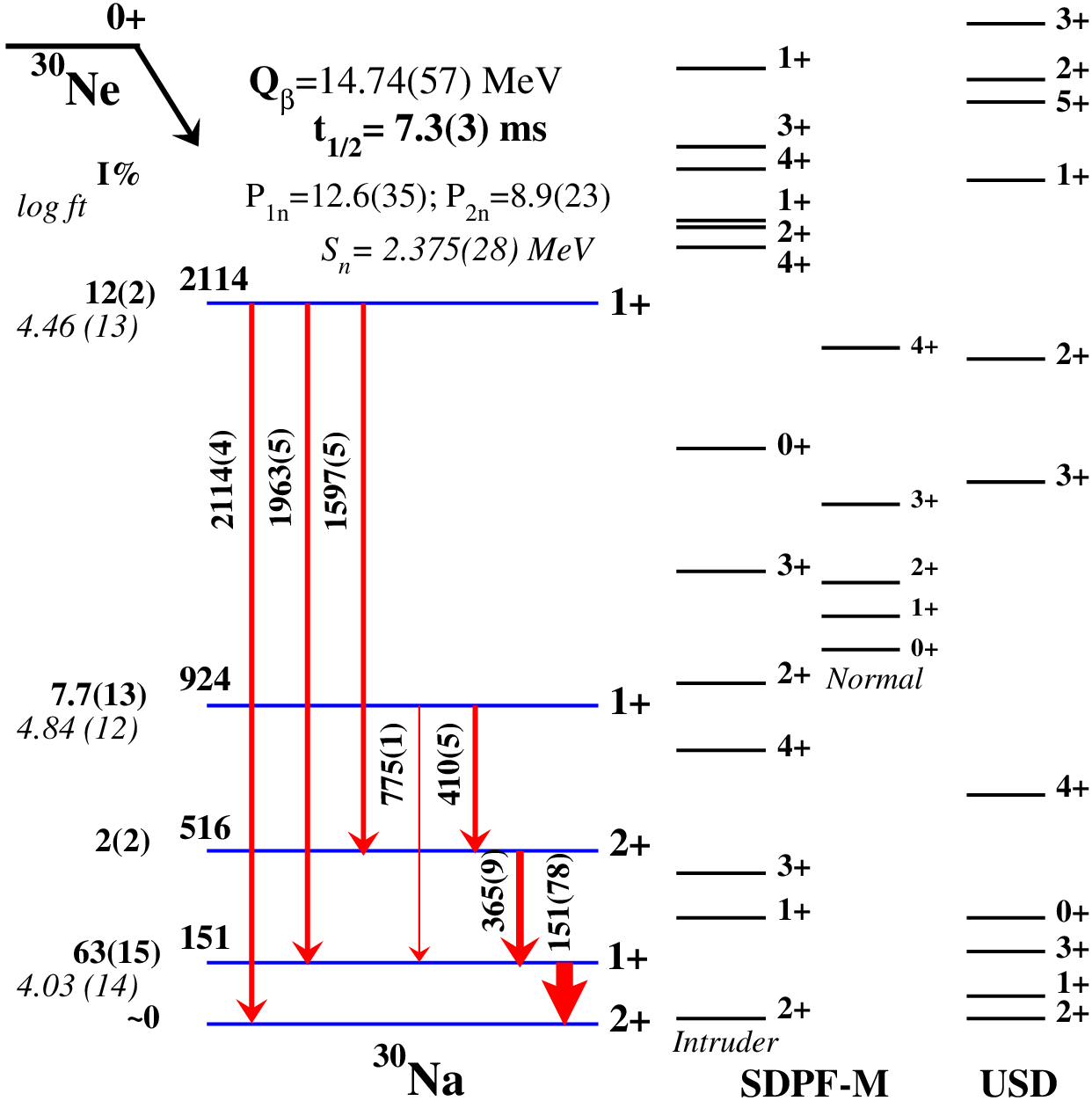}
\caption{(Color online) Experimental and theoretical level schemes for
$^{30}$Na  (For MCSM,  levels  from  0$^+$ -  4$^+$  only are  shown).
Energies  are   in  keV.    Absolute  intensities  for   the  $\gamma$
transitions  (\%) are  indicated.  The  ground state  of  $^{30}$Na is
$2^+$ \cite{huber} consistent with negligible beta branching.}
\label{fig3}
\end{center}
\end{figure}

The  $\gamma$  rays observed  up  to 50  ms  after  implantation of  a
$^{30}$Ne  ion,   correlated  with  a   decay  event,  are   shown  in
Fig. \ref{fig1}.  Seven $\gamma$ lines are identified to correspond to
transitions in $^{30}$Na, indicated in Fig.  \ref{fig1}.  Only the 151
keV had been  reported before \cite{reed}.  All but  the 2114 keV line
are  in coincidence  with the  151  keV (see  Fig. \ref{fig2}),  which
satisfies the energy  sum rule.  The 365- and  410 keV transitions are
seen  to be  in mutual  coincidence and  coincident with  the  151 keV
$\gamma$ line, which along  with the coincidences observed between the
365 keV and 1597 keV transitions, implies four excited states at 151-,
516-, 924- and  2114 keV.  The 2114 keV level  is further supported by
its direct decay  to the ground state as well  as coincidences to show
its    decay    to   the    151    keV    state.     Based   on    the
fragment-$\beta$-$\gamma$  -$\gamma$ coincidences  and the  energy and
intensity  sum rules,  the first  level scheme  of $^{30}$Na  has been
constructed    following   the    $\beta^-$    decay   of    $^{30}$Ne
(Fig. \ref{fig3}).

\begin{figure}
\begin{center}
\includegraphics[scale=0.28]{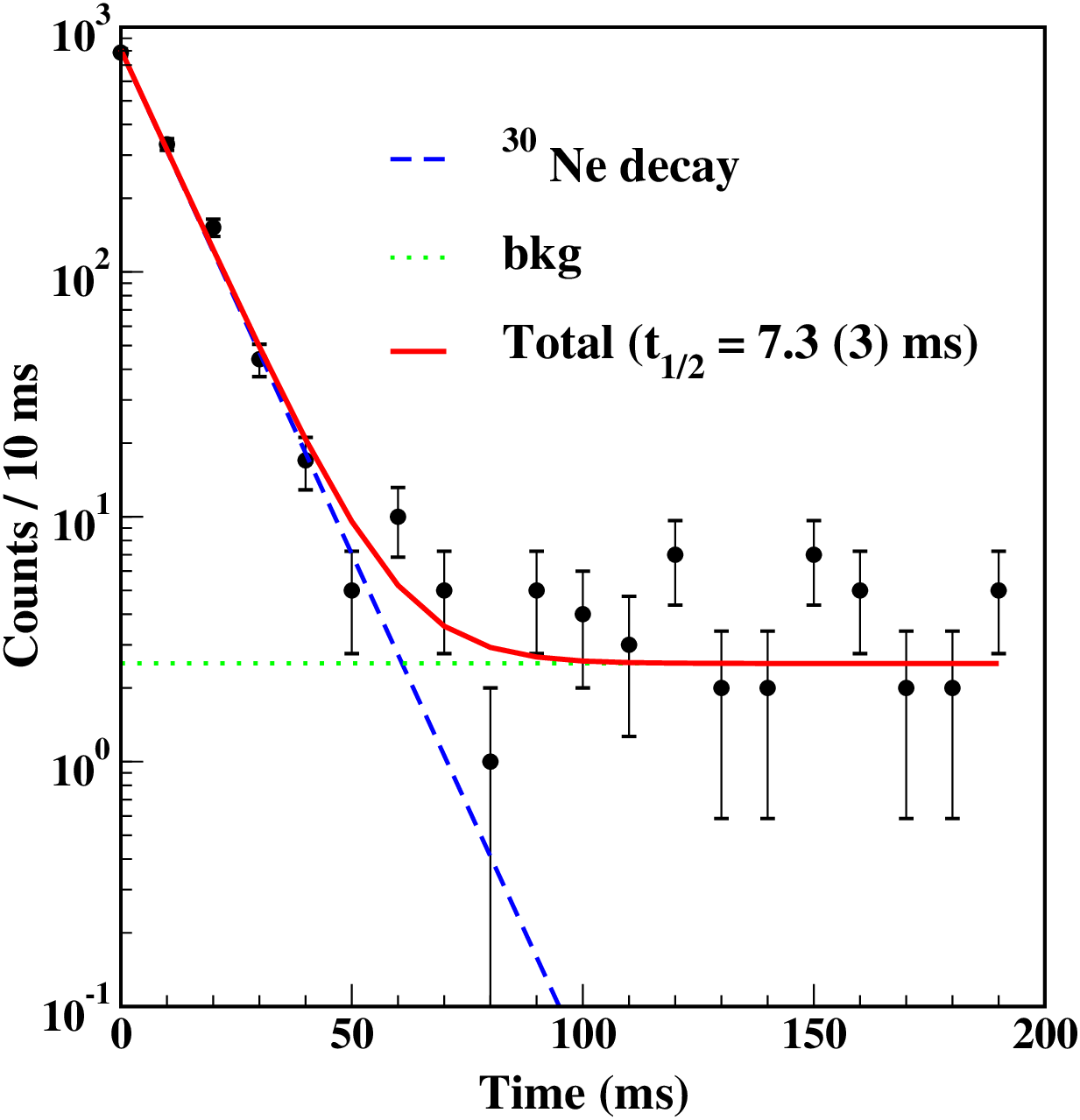}
\caption{(Color   online)  Time   spectra  for   $^{30}$Ne   decay  in
coincidence  with  the  151  keV  $\gamma$,  fitted  with  a  decaying
exponential  and a  constant background  to extract  the  half-life of
$^{30}$Ne.}
\label{fig4}
\end{center}
\end{figure}

The  absolute  intensities  of  the  bound  levels  populated  in  the
$\beta^-$ decay were calculated using the measured SeGA efficiency and
the total  number of $^{30}$Ne  decay events, 127({\it 14})  x 10$^2$,
obtained from  the intensities  of $\gamma$ transitions  in $^{30}$Mg,
consistent with that obtained from a fit to the decay curve. The $P_n$
and $P_{2n}$, expected to be  significant in neutron rich nuclei, were
estimated  to be  12.6({\it 35})\%  and 8.9({\it  23})\% respectively,
from  the intensities  of transitions  in the  grand  daughter nuclei,
$^{30}$Mg, $^{29}$Mg, and $^{28}$Mg, populated in $0n$, $1n$, and $2n$
emission.  This is consistent with the adopted value of $< 26\%$ \cite
{nndc}.  The decay curve in coincidence with the 151 keV transition in
$^{30}$Na was also used to  extract the decay half life. The half life
obtained is 7.3({\it  3}) ms (see Fig. \ref{fig4}),  in agreement with
the adopted value, 7({\it 2})  ms \cite{nndc}. The log {\it f}t values
for the observed states were calculated from the absolute intensities,
the  measured  half-life  and  the  $Q_{\beta^-}$  value  \cite{audi},
according  to  Ref.  \cite   {logft}  (ignoring  the  weak  unobserved
transitions) and are listed in Table 1.

Allowed $\beta$  transitions from the 0$^+$ ground  state of $^{30}$Ne
will populate  only $1^+$  states in the  daughter $^{30}$Na.  The log
{\it f}t values for the  $\beta$-decay branches to the 151-, 924-, and
2114  keV states  (4.03 to  4.84) imply  allowed  $\beta$ transitions.
Thus a firm assignment of $J^\pi = 1^+$ is made to these three states.
The  ground  state of  $^{30}$Ne,  with $N=20$,  is  known  to have  a
dominance of $fp$ intruder configurations \cite{Yana}. The most likely
$\beta^-$ decay scenario from such a $2p2h$ state is the conversion of
one  $0d_{3/2}$ neutron into  a $0d_{5/2}$  proton, creating  a $2p2h$
state  in $^{30}$Na.   This would  lead to  stronger  $\beta$ branches
(lower  log  {\it f}t  )  to  the  intruder-dominant $1^+$  states  in
$^{30}$Na.   The observed  strong branches  to the  151- and  2114 keV
states   thus  demonstrates  their   intruder  dominance,   while  the
relatively weaker  branch to the 924  keV state suggests  a purer $sd$
structure of this state.

Shell model  calculations carried out in  the {\it sd}  shell with the
Universal (USD) interaction \cite {alex} predict only two $1^+$ states
below 3  MeV at 66 and 2511  keV.  This along with  the discrepancy in
predicting  the quadrupole moment  of the  ground state  of $^{30}$Na,
highlights the limitation of the pure $sd$ model space for this $N=19$
nucleus.  The measured quadrupole moment of $^{30}$Na \cite {keim}, is
reproduced by the MCSM  calculations \cite {utsuno_na} predicting 98\%
$2p2h$ configuration of the  ground state.  Hence states with intruder
character at low excitation energies are expected in $^{30}$Na.

The MCSM  calculations with the SDPF-M  interaction \cite {utsuno_na},
which gives a narrow $N=20$ shell gap (3.3 MeV) for Na, were performed
in the $sd-p_{3/2}f_{7/2}$ space. These calculations incoporate mixing
between all possible configurations.  The excited states of $^{30}$Na,
their  B(GT) values  (no  quenching assumed)  and the  electromagnetic
transition strengths between the states were obtained.

\begin{table*}
\caption{Excitation  energies,  log{\it   ft}  values  and
$\gamma$-branching  ratios for the observed levels in
$^{30}$Na, and the predictions of the MCSM calculation with the
SDPF-M  interaction \cite{utsuno_na}.   For MCSM,  the  probability of
2p2h contribution is also indicated.}
\begin{minipage}{\textwidth}
\begin{tabular}{ccccccccc}
\hline
\hline
 &  {\bf Exp. } & & & & & & {\bf SDPF-M} & \\ 
 {\bf $E_x$ (keV) }           & {\bf $J^\pi$}  & {\bf log {\it f}t}       & 
 {\bf $\gamma$-branching(\%)} & ~~{\bf $E_x$ (keV) } & {\bf $J^\pi$}      & 
 {\bf log {\it f}t}\footnotemark & {\bf $\gamma$-branching(\%)}\footnotemark & 
 {\bf 2p2h(\%)} \\
\hline
151({\it 1})  & ~$1^+$ & 4.03({\it 14}) & 0(100)                 & 
~310          & ~$1^+$ & 3.9            & 0(100)                 & 98 \\
516({\it 1})  & ~$2^+$ & -              & 151(100)               & 
~980          & ~$2^+$ & -              & 310(99.8)              & 87 \\
924({\it 1})  & ~$1^+$ & 4.84({\it 12}) & 516(83); 151(17)        & 
~1210         & ~$1^+$ & 4.9            & 980(64); 310(32); 0(4)   & 23 \\
2114({\it 2}) & ~$1^+$ & 4.46({\it 13}) & ~~516(36); 151(36); 0(28)  & 
~2380         & ~$1^+$ & 4.9   &   1210(1); 980(7); 310(88); 0(4.5) & 98 \\
-             &  -     & -     &             -                   & 
~2820         & ~$1^+$ & 4.2   & 1210(0.1); 980(36); 310(40); 0(24) & 96 \\
\hline
\hline
\end{tabular}
\footnotetext{ground  state of $^{30}$Ne  had 4\%  of $0p0h$,  74\% of
$2p2h$  and 22\%  of $4p4h$  configuration} \footnotetext{$\gamma$-ray
energies taken from experimental data}
\end{minipage}
\label{table1}
\end{table*}

The MCSM calculations predict four  bound $1^+$ states at 310-, 1210-,
2380-, and  2820 keV (Table 1  and Fig.  \ref{fig3}).   The lowest two
calculated  $1^+$ states,  though located  higher in  energy  than the
experimental ones at 151 keV and 924 keV, correspond well in their log
{\it f}t values  and $\gamma$ decay.  The experimental  2114 keV state
agrees better in energy with the  2380 keV level, but its log {\it f}t
value and $\gamma$ decay branches correspond to those of the predicted
$1^+$   state  at   2820  keV.    In  the   latter  and   more  likely
identification,   the  experimental   non  observation   of   a  state
corresponding to 2380  keV would result from its  $\sim$5 times weaker
population than the 2820 keV implied by the larger calculated log {\it
f}t  value. The 516  keV state  is not  directly populated  by $\beta$
decay, excluding  a $1^+$ assignment. The  decay of the  516 keV state
only to  the $1^+$ state, excludes  J$^\pi =3^+$, as it  would favor a
pure low-energy E2 over a  higher energy M1 decay.  Hence the possible
candidate is  the SDPF-M  $2^+$ state at  980 keV, predicted  to decay
almost 100\%  to the  lowest $1^+$  level, as does  the 516  keV state
(Table 1).

Prior studies of  $^{30}$Na, by intermediate-energy Coulomb excitation
at the  NSCL \cite  {prity} and the  (p,p$^\prime$) reaction  at RIKEN
\cite{elekes}  measured  $\gamma$  rays  of 433(16)  and  403(18)  keV
respectively. Though close  in energy to the 410  keV from the present
work, they are  unlikely to represent the same  transition as it would
require a multi-step  decay process involving the 365  keV and 151 keV
transitions, not  seen in Refs.  \cite {prity, elekes}.  The predicted
$3^+$   SDPF-M   state   at   430   keV  remains   the   most   likely
identification. The 360(13) keV line observed in neutron knockout from
$^{31}$Na \cite{elekes} could correspond  to the 365 keV line reported
here.

\begin{figure}[b]
\begin{center}
\includegraphics[scale=0.335]{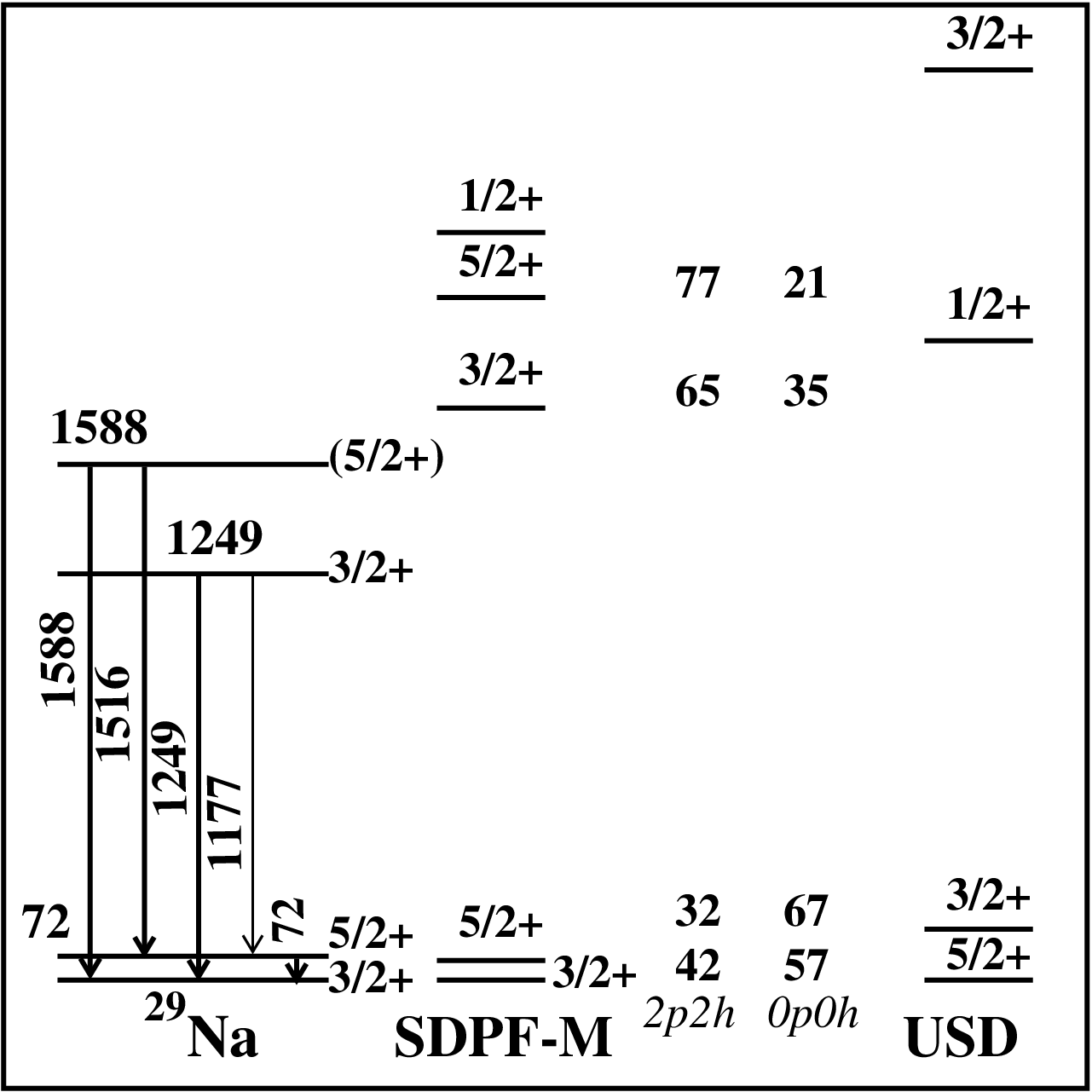}
\caption{     Partial    level     scheme     of    $^{29}$Na     from
Ref.  \cite{tripathi}.  The  $J^\pi$  assignments  to  the  states  at
$\sim$1.5 MeV are from the  present work.  Configuration of the levels
($2p2h$ and  $0p0h$) obtained with the SDPF-M interaction are 
listed (in \%).}
\label{fig5}
\end{center}
\end{figure}  

An  analysis of the  wave functions  of the  predicted levels  in MCSM
calculations  reveals that  the  second  $1^+$ state  at  1210 keV  is
dominated  by $0p0h$  configurations,  whereas the  other three  $1^+$
states have  almost pure  intruder configurations, mainly  $2p2h$ with
$\sim$  1\%  $4p4h$ (see  Table  1).   This  fits perfectly  with  the
experimental picture,  the 151  keV and 2114  keV states  with smaller
log{\it f}t values  as dominant $2p2h$ $1^+$ states  while the 924 keV
state with a smaller branch as the dominant $0p0h$.  The prediction of
no strong  connecting transitions between 2114  keV - 924  keV and 924
keV - 151  keV states, corroborated well by  experimental data further
projects  their  different character.   Thus  a  clear `inversion'  is
observed, the first excited  state with dominant $sd$ configuration is
at 924 keV, lying above  many intruder dominated states.  The location
of this `normal' excited state,  observed for the first time in exotic
Na isotopes,  is extremely important to determine  exactly the $sd-fp$
shell gap. The ground state properties, on the other hand, can provide
only the upper limit.

The  situation  is  different  for  the less  exotic  $^{29}$Na  \cite
{tripathi},  where  the  intruder  dominated states  occur  at  higher
excitation energy, as  seen from Fig. \ref{fig5}.  The  1249 keV level
is  assigned a J$^\pi  = 3/2^+$  from the  present work,  assuming its
population in  $\beta$-n emission  (Fig.  \ref{fig1}) from  an unbound
$1^+$ in  $^{30}$Na by a $l=0$  neutron (most probable due  to the low
energy).  This corresponds to the SDPF-M  state at 1760 keV and is the
first   excited  state   with  dominant   intruder   configuration  in
$^{29}$Na. The  comparison of  the lowest excitations  in $^{29,30}$Na
thus illustrates  the mechanism of  intrusion, {\it i.e},  states with
dominant intruder configuration moving to lower excitation energies by
gaining   correlation  energy,   as  the   neutron   number  increases
\cite{utsuno_na}.

To  recapitulate,  a  comparison  of  the  excited  states,  weak  and
electromagnetic branching  ratios in the $\beta^-$  decay of $^{30}$Ne
to $^{30}$Na  with shell model predictions  in the $sd-f_{7/2}p_{3/2}$
space, clearly  demonstrates the `inversion',  {\it i.e.} a  number of
{\it  intruder dominated}  states  lie below  the  lowest {\it  normal
dominant} state.   The decay branches agree  surprisingly well, though
the calculations tend to over-predict the excitation energies, a trend
also  seen in  $^{29}$Na.  For  the first  time, excited  `normal' and
`intruder'  states  have  been  unambiguously  identified  inside  the
`island  of  inversion'.  Their  relative  position provides  valuable
information for  better determining the $d_{3/2}$ -  $f_{7/2}$ gap and
thus the evolution of shell structure with isospin.

The authors appreciate the NSCL operations staff for their help.  This
work was supported by the NSF grants PHY-01-39950 and PHY-06-06007 and
in part  by a Grant-in-Aid for Specially  Promoted Research (13002001)
from the MEXT of Japan.

\end{document}